\documentclass[aps,pra,reprint,superscriptaddress,amsmath,amsfonts,amssymb,longbibliography]{revtex4-1}

\usepackage[utf8]{inputenc}
\usepackage{graphicx} 
\usepackage{float} 
\usepackage{xcolor}

\usepackage{amssymb}
\usepackage{dsfont}

\usepackage{physics}

\usepackage{listings} 

\usepackage{color}

\usepackage{hyperref}

\usepackage[nodisplayskipstretch]{setspace}
\usepackage{hyperref}% add hypertext capabilities

\begin{document}

\title{Classical approach to equilibrium of out-of-time ordered correlators in mixed systems}

\author{Tomás Notenson}
\affiliation{Departamento de F\'isica ``J. J. Giambiagi'' and IFIBA, FCEyN,
Universidad de Buenos Aires, 1428 Buenos Aires, Argentina}
\author{Ignacio Garc\'ia-Mata}
\affiliation{Instituto de Investigaciones Físicas de Mar del Plata (IFIMAR), Facultad de Ciencias Exactas y Naturales,
Universidad Nacional de Mar del Plata, CONICET, 7600 Mar del Plata, Argentina}
\author{Augusto J. Roncaglia}
\affiliation{Departamento de F\'isica ``J. J. Giambiagi'' and IFIBA, FCEyN,
Universidad de Buenos Aires, 1428 Buenos Aires, Argentina}
\author{Diego A. Wisniacki}
\affiliation{Departamento de F\'isica ``J. J. Giambiagi'' and IFIBA, FCEyN,
Universidad de Buenos Aires, 1428 Buenos Aires, Argentina}

\date{March 11, 2023}

\begin{abstract}
The out-of-time ordered correlator (OTOC) is a measure of scrambling of quantum information. Scrambling is intuitively considered to be a significant feature of chaotic systems and thus the OTOC is widely used as a measure of chaos. For short times exponential growth is related to the classical Lyapunov exponent, sometimes known as the butterfly effect. At long times the OTOC attains an average equilibrium value with possible oscillations. For fully chaotic systems the approach to the asymptotic regime is exponential with a rate given by the classical Ruelle-Pollicott resonances. In this work, we extend this notion
{to the more generic case of systems with mixed dynamics, in particular using  the standard map and we are able to show that the relaxation to equilibrium of the OTOC is governed by generalized classical resonances.}  
% by showing that classical generalized resonances govern the relaxation to equilibrium of the OTOC in the ubiquitous case of a system with mixed dynamics, in particular, the  standard map.
% 
\end{abstract}

\maketitle
\section{\label{sec:introduction} Introduction}
The out-of-time ordered correlator (OTOC) has been proposed as a measure of spreading of information and scrambling \cite{larkin1969quasiclassical,maldacena2016bound}. Recently, it was noticed that for chaotic systems the short-time behavior is characterized by an exponential growth which can be directly related (in some cases) to the classical Lyapunov exponent. A straightforward generalization would lead to characterizing systems with exponential growth as chaotic and naming the growth rate  ``quantum Lyapunov exponent''. However this is not a universal behavior. There are systems that are not chaotic where the OTOC grows exponentially  \cite{ali2020chaos,hashimoto2020exponential,morita2021extracting,HummelPRL2019,XuPRL2020,RozenbaumPRL2020,kidd2021saddle} and the converse is also true, there are systems universally accepted as quantum chaotic where the growth is much slower \cite{kukuljan2017,Motrunich2018,PhysRevB.99.054205,PhysRevE.100.042201,Shukla2022,moessner2017}. 

Interestingly, a lot of information about the dynamics, and some underlying classical features, can be obtained  at later times. After the initial growth of the OTOC, the system approaches an equilibrium state with oscillations. The study of both the equilibrium value \cite{hashimoto2017,JGMW2018,rammensee2018,huang2019finite,sinha2021fingerprint,Markovic2022} and the oscillations \cite{PhysRevE.100.042201,fortes2020signatures}  has allowed us to distinguish chaotic from regular behavior.
However, evaluation of equilibrium-based measures requires evolving up to long times and computing  averages over large time windows. This can be experimentally unattainable. 
Nevertheless, at a much shorter time scales there is another classical quantity that can be used to characterize the chaotic nature of a system. In a simplified way, fully chaotic systems present two main features: exponential separation of initial conditions (Lyapunov regime), and mixing. The mixing property implies exponential decay of correlation functions which is characterized by the largest Ruelle-Pollicott resonance. The Ruelle-Pollicott resonances are the point spectrum of the Perron-Frobenius operator expressed in a suitably defined functional space~\cite{Blank2002}. For fully chaotic quantum maps on the torus, it was shown~\cite{PhysRevLett.121.210601} that the OTOC approaches equilibrium exponentially, and the rate is given by the classical  Ruelle-Pollicott resonance with the largest modulus.

The case of systems with mixed dynamics (i.e. neither fully integrable nor fully chaotic), although much more generic, is less studied. In such systems the  Ruelle-Pollicott resonances are not strictly well defined. However, through approximation schemes, the spectrum of the Perron-Frobenius can be obtained. In this case, it is observed that there are eigenvalues that tend to the unit circle, associated with regular behavior, and there are other eigenvalues that persist inside the unit circle, and are considered generalized resonances. The eigenfunctions of these resonances are located in chaotic regions in phase space~\cite{Manderfeld2001,Weber2001,KhodasFishmanAgam,fishman2002relaxation,nonnenmacher2003spectral}.  

In this work, we use numerical methods to obtain the generalized resonances and establish a correspondence with the rate of approach to equilibrium of the OTOC, given by the decay of a four point correlator. The numerical calculations are done on the standard map \cite{Chirikov1979} which can be parametrically tuned from integrable to chaotic. The values obtained for the resonances reproduce very well the structure in the decay rate as a function of the chaos parameter. Interestingly, we also find that in this case, the quantum evolution is able to reveal classical related effects such as the appearance and disappearance of stable islands in the classical phase space. This is clearly reflected in the rich structure of the decay rates as a function of the chaos parameter.

This paper is organized as follows. In Sec.~\ref{sec:O1 in maps} we define the relevant quantities and observables to be used. In Sec.~\ref{sec:standard} we compute numerically the decay rates as a function of the chaos parameter for the standard map.
Then, in Sec.~\ref{sec:sepctrumPF}, we describe the numerical method to compute the spectrum of the Perron-Frobenius operator. Once we have established the method to compute the generalized resonances, we show their relationship to the decay rates in Sec.~\ref{sec:relationship}, leaving Sec.~\ref{sec:conclusions} for the concluding remarks.

\section{\label{sec:O1 in maps} Out-of-time-ordered correlators in quantum maps}
Generically, the OTOCs are defined as
\begin{equation}
C(t)=\langle |[{W}(t),{V}]|^2\rangle
\end{equation}
where $V$ and $W$ are two operators, with $W(t)$ being the Heisenberg operator at time $t$, and the average is taken with respect to 
{an equidistribution of states, which can be viewed as the infinite temperature state for Hamiltonian systems.}
%a thermal state (here we consider the infinite temperature case). 
This quantity was introduced by Larkin and Ovchinnikov to study semiclassical approaches to superconductivity \cite{larkin1969quasiclassical}, and received considerable attention later due to its connection to the scrambling of quantum information 
chaotic systems and the dynamics of black holes \cite{shenker2014black,shenker2014multiple,shenker2015stringy}. 
In fact, when the operators have local support in different regions of a many body system, the spreading of information can be quantified by the OTOC.
They are called OTOCs, mainly because their expansion as $C(t) = -2(O_1 - O_2)$ contains the term
\begin{equation}
 O_1(t)=\langle {W}(t){V}{W}(t){V} \rangle
\end{equation}
which is  an out-of-time-ordered product, and another correlator $O_2 = \langle {W}(t)^2{V}^2 \rangle$. 

\begin{figure}
\includegraphics[width=.43\textwidth]{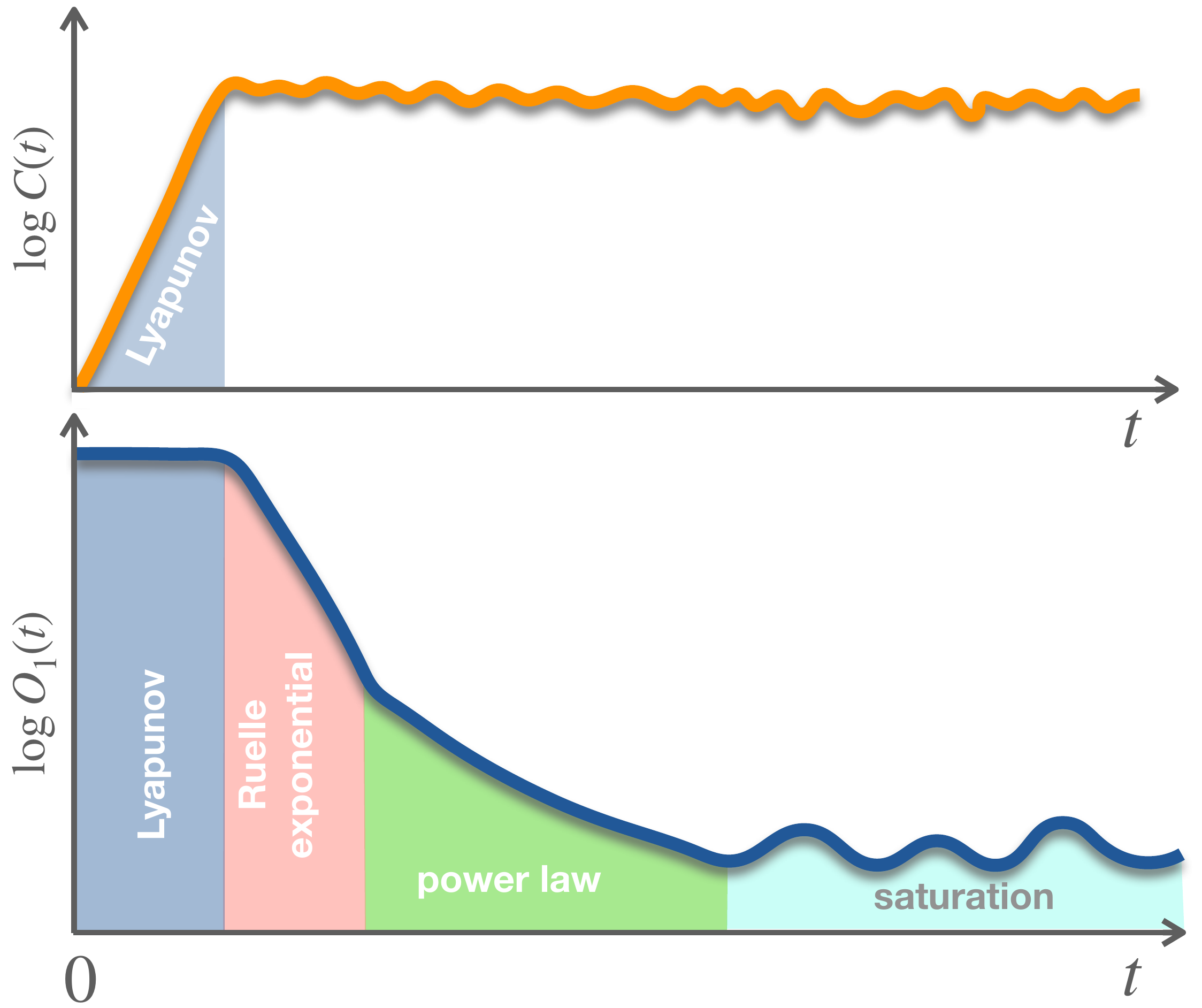}%{fig1tomy.pdf} %
\caption{\label{fig:analysis} 
Sketch of the typical time dependence of $O_1 = \langle {W}(t){V}{W}(t){V} \rangle$ for mixed quantum maps exhibiting different
behaviors in the short-, intermediate-, and long-time regimes. At short times, namely for times below the Ehrenfest time $t<t_E$, it has an approximate constant value (blue shading). We name this short-time behavior the Lyapunov regime because of the exponential growth in $C(t) = \langle |[{W}(t),{V}] |^2\rangle = -2(O_1 - O_2)$ with $O_2 = \langle {W}(t)^2{V}^2 \rangle$ (see the top panel and \cite{PhysRevLett.121.210601} for more detail). Next, the correlator has an exponential decay related to the Ruelle resonances of the classical map (red shading). The succeeding regime is the power-law regime, which we associate with regular vestiges of the map (green shading). For long times, $O_1$ oscillates around his saturation value (cyan shading).}
\end{figure}

The early time dynamics of the OTOCs in quantum systems with a classical chaotic counterpart is rather universal, and can be characterized by exponential growth with time and saturation, as depicted in the schematic in Fig.~\ref{fig:analysis}(a). This growth shows a clear relationship with the classical Lyapunov exponent, and it lasts up to the so-called scrambling time at which this quantity saturates \cite{PhysRevLett.118.086801,PhysRevLett.121.210601,chavez2019quantum,sinha2021fingerprint,hashimoto2020cho,PhysRevE.99.012201}.  Thus, it is remarkable that the short-time dynamics of the OTOCs contains a signature of the classical behavior. For one-body systems the saturation time is proportional to the Ehrenfest time \cite{PhysRevLett.118.086801,PhysRevLett.121.210601} {($\tau_{\rm E}\sim \ln D/\lambda$, where $D$ is the Hilbert space size).} 

After the Lyapunov regime, the approach to equilibrium of $C(t)$ is governed by the correlator $O_1(t)$. For fully chaotic systems this decay is exponential, and it was hinted \cite{polchinski2015}, and demonstrated for chaotic quantum maps~\cite{PhysRevLett.121.210601}, to be related to classical quantities  known as Ruelle-Pollicott resonances \cite{pollicott1985rate,ruelle1986,ruelle1987resonances}. The Ruelle-Pollicott resonances can be loosely defined as the point spectrum of the evolution operator of classical distribution functions in the case of fully chaotic systems.
A classical dynamical system evolves through the one step evolution operator defined by\begin{equation}\label{eq:perron_frobenius}
    \rho(x,t) = \mathcal{L}^t \rho(x,0) = \int_{\mathcal{M}}\!dx_0 \,\delta[x-f^t(x_0)] \rho(x_0,0),
\end{equation}
where $\rho(x,t)$ is the density of representative points at $x$ in the phase space at time $t$ in the zero coarse grained limit, $\mathcal{L}$ is the Perron-Frobenius (PF) evolution operator, $f(\cdot)$ is the map and the integral runs throughout phase space $\mathcal{M}$ ~\cite{cvitanovic2005chaos}. When defined in $\mathbb{L}^2$ the PF is unitary. However this is rather restrictive. A way to unveil the spectrum of the PF operator is to allow for distributions. Then one can define a functional space adapted to the dynamics, smooth along the unstable direction and (possibly) singular along the stable dimension. For uniformly hyperbolic, chaotic systems, the spectrum of the thus expressed PF operator consists of a point spectrum and an  essential spectrum~\cite{reed1980methods}. The essential spectrum is  bounded by a small radius $r<1$. The eigenvalues composing the point spectrum are the Ruelle-Pollicott resonances $\{\lambda_i\}$, where $r\leq |\lambda_i|\leq 1$~\cite{Blank2002}. The case $\lambda_0 = 1$ corresponds to the invariant density. The correlation function decay is then determined by resonances such that $1>|\lambda_1|>|\lambda_2|>\ldots$, and typically, if there are no degeneracies, it is dominated by $\lambda_1$. For quantum maps on the torus in the fully chaotic regime it has been numerically proven~\cite{PhysRevLett.121.210601} that the decay of $O_1$ is given by 
\begin{equation}
\label{eq:o1lambda1}
    O_1(t)\sim |\lambda_1|^{2t}.
\end{equation}
{The theoretical background for this behavior relies on the spectral correspondence of coarse grained propagators of densities, from quantum to classical, in suitably  limits (where the order of limits matters) of zero coarse graining and a vanishing effective Planck constant~\cite{nonnenmacher2003spectral}. Once the correspondence is established, a finite gap $(1-|\lambda_1|)$, typical of chaotic behavior, leads to decay like Eq.~\eqref{eq:o1lambda1}. For quantum systems with dissipation it has been shown that the OTOC $C(t)$, after the initial Lyapunov growth, decays as Eq.~\eqref{eq:o1lambda1}\cite{PhysRevLett.121.210601,Bergamasco2022}. } 

The spectrum in the case of mixed phase space dynamics is neither as simple to describe nor to compute, and studies about it do not abound. In \cite{khodas2000relaxation,KhodasFishmanAgam,Manderfeld2001,Weber2001} using coarse graining and truncation schemes, a description of the spectrum is made in which eigenvalues that are close to the unit circle are identified with regular island regions and some persistent (``frozen'') eigenvalues, with a modulus strictly smaller than 1, can be identified as corresponding to densities located in the chaotic regions of phase space. The latter are  ``generalized'' Ruelle resonances.

We will show below that for a system with mixed dynamics,
the approach to equilibrium displays a clear exponential regime determined by these so-called generalized resonance. This is the regime that we study in the paper. We have also observed a second power-law regime that is far less studied, that lies outside the scope of this work, and we leave it for future endeavors.

Let us now define the relevant quantities that we will consider in this work. The numerical calculations will be done on quantum maps on the 2-torus, more specifically, the Chirikov standard map~\cite{Chirikov1979,DimaScholar}. After quantization of the two-dimensional phase space, the periodic properties impose discreteness on the position and momentum, and the effective Planck constant is related to the dimension of Hilbert space $D$ as $\hbar_{\rm eff}=1/(2\pi D)$. The operators we use to evaluate the OTOCs in quantum maps are the so-called position and momentum operators
\begin{equation}
    {X} \equiv \frac{{U}_S - {U}^\dagger_S}{2i}\text{, } \, \, {P} \equiv \frac{{V}_S - 
    {V}^\dagger_S}{2i},
\end{equation}
which are Hermitian and defined in terms of the unitary Schwinger shift operators $U_S$ and  $V_S$ \cite{schwinger1960unitary}: 
\begin{equation}
    {V}_S =\sum\limits_q |q + 1 \rangle \langle q| \text{, } \, \, {U}_S = \sum\limits_q e^{2 \pi i q/D} |q\rangle \langle q |
\end{equation}
where $|q\rangle$, and $|p\rangle$ are, respectively, position and momentum states for a $D$ dimensional system with periodic boundary conditions (with $q, p = 0,..., D - 1$). They are related by the discrete Fourier transform satisfying  $\langle p | q \rangle = e^{-2\pi iqp/D}$  and ${V}_S|D-1\rangle = |0\rangle$. In the semiclassical limit ${X}$ and ${P}$ approximate the position and momentum operators, respectively.
Thus,  for the OTOC we take the average over the maximally mixed state $\rho = \mathds{1}/D$:
\begin{equation}
    O_1(t) = \langle {P}(t){X}{P}(t){X} \rangle = \text{Tr}({P}(t){X}{P}(t){X})/D
\end{equation}
where for a quantum stroboscopic map that evolves with the unitary operator $U$, ${X}(t) = ({U}^\dagger)^t {X} {U}^t$.
{For the numerical calculations in the following sections, all the operators are represented as $D\times D$ matrices and $O_1(t)$ is obtained through successive matrix  multiplications.}

\section{Decay of $O_1$ for the standard map}
\label{sec:standard}
\par Mixed systems are ubiquitous within Hamiltonian dynamics, however, they have been much less studied than completely chaotic and integrable systems. This is because the coexistence of different dynamics produces very complex structures. In this work, we want to understand the effect that chaotic regions and islands of stability have on the decay of $O_1$. For this reason, we have focused on the standard map, a paradigmatic model within both classical and quantum studies. This is generated by the time-dependent Hamiltonian 
\begin{equation} 
H(q,p,t)=p^{2}/2+K/(2 \pi)^{2} \cos(2 \pi q)\sum_j \delta(t-j)
\end{equation}
where $K$ is the strength of $\delta$ kicks. Due to the periodicity of $\cos(\cdot)$ the dynamics can be considered on a cylinder (by taking $p \text{ mod } 1$) or on a torus (by taking both $p,q \text{ mod } 1$) \cite{DimaScholar}. We are interested in the second description. So, the corresponding classical map is,
\begin{equation} 
\label{eq:clstdmap}
    \left. \begin{array}{ll}
             p_{n+1} &= p_n + \frac{K}{2\pi}\sin(2\pi q_n) \\
             q_{n+1} &= q_n + p_{n+1}
             \end{array}
   \right\}\text{mod }1
\end{equation}

\noindent  
\begin{figure}[h]
\includegraphics[width=0.5\textwidth]{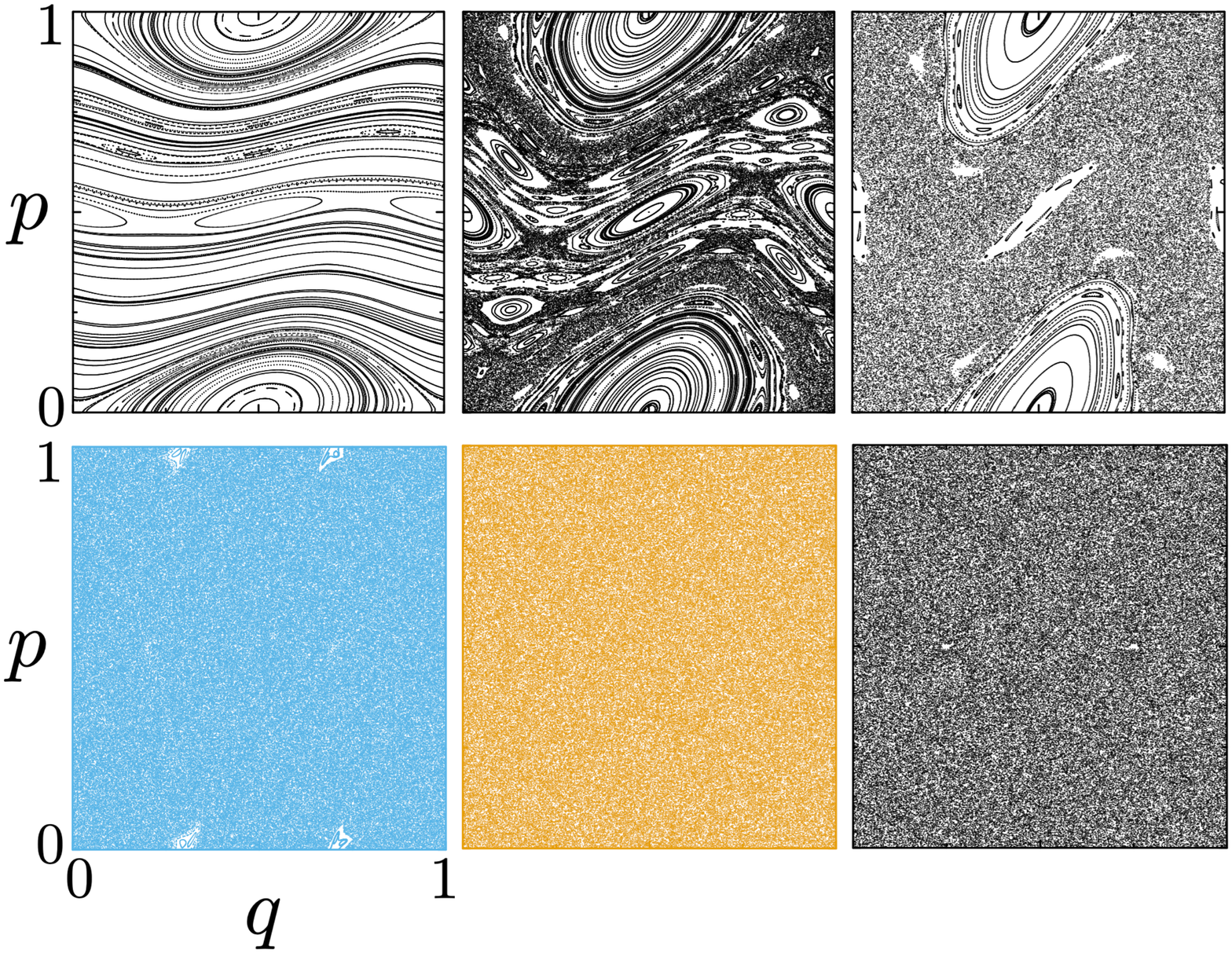} %%%{fig2_cledf2.pdf}
\caption{\label{fig:edf} Sample phase spaces for the standard map Eq.~(\ref{eq:clstdmap}) and $K=0.4,\, 1,\, 1.8,\,6.6,\,17,\, 18.86$ (from left to right, top to bottom). The labels are the same for every panel. The colors in the  bottom-left and middle panels are meant to correspond to the $K$ values used in Fig.~\ref{fig:numeric_decay}. Notice that for very large $K$ (bottom right panel) small regular islands can be seen.}
\end{figure}

The classical version of this model manifests a regular to chaotic transition (as a function of driving strength $K$), enabling us to benchmark the behavior of the OTOC against the presence or absence of classical chaos and islands of stability.
For small values of $K$ the dynamics is regular. Below a certain critical value $K_c$, the motion in momentum is restricted by the Kolmogorov-Arnold-Moser (KAM) curves. These are invariant curves with an irrational winding number that represent quasi-periodic motion, and they are the most robust orbits under nonlinear perturbations \cite{liebermann1983irregular}. At $K_c = 0.971635406 ...$, the last KAM curve, with the most irrational winding number, breaks down \cite{greene1979method}. Above $K_c$, there is unbounded diffusion in $p$. This critical point is important for the map on the cylinder, non-periodic in $p$, which is not considered in this work. For very large values of $K$ the motion is essentially chaotic with no visible islands. However, as $K$ is increased, at approximately periodic $K$ values, small islands do appear, and then disappear. An illustration of  this behavior is presented is presented in Fig.~\ref{fig:edf}, where we see that there are no islands for large $K=17$ and then at $K=18.86$ two visible islands reappear. 

\par The quantum standard map is defined by the Floquet evolution operator,
\begin{equation}
    {U}(K) = e^{- \frac{i}{\hbar_{\rm eff}} {P}^2/2}e^{- \frac{i}{\hbar_{\rm eff}} \frac{K}{4\pi^2} \cos(2\pi {X})}
\end{equation}
and its dynamics is given by a sequence of free propagations interleaved with periodic kicks. The advantage of the previous formulation is that the numerical implementation of the time evolution and the corresponding diagonalization becomes very efficient using fast Fourier transformations \cite{ketzmerick1999efficient}. For the map we consider, the kick strength is the chaos parameter.

\begin{figure}[ht]
    \includegraphics[width=0.95\linewidth]{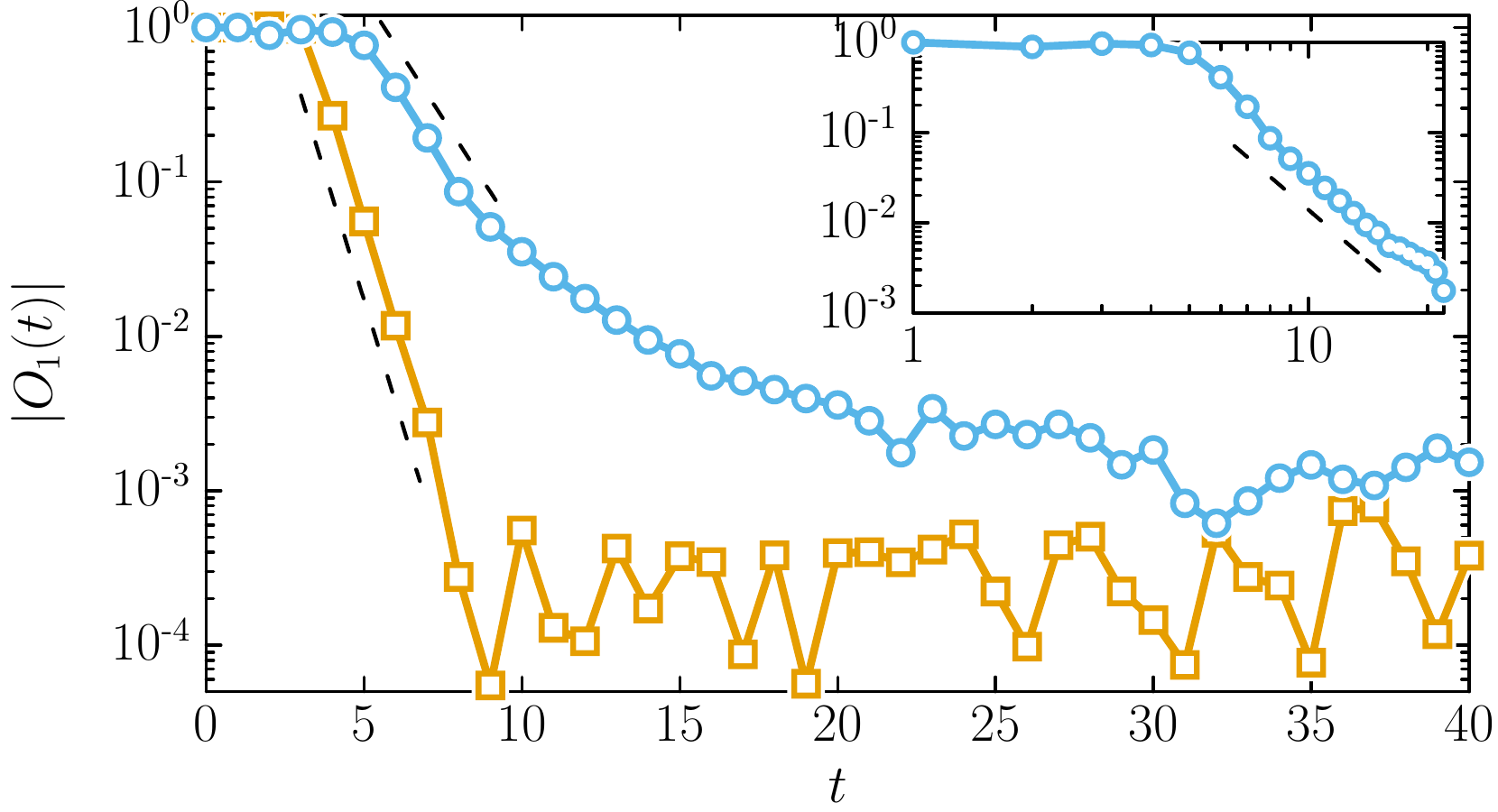}
\caption{\label{fig:numeric_decay} 
$|O_1|$ versus time for $K=6.6$ (circles) and $K=17$ (squares). These two $K$ values correspond to the bottom left and middle panels in Fig.~\ref{fig:edf}. For $K=6.6$, we observe an exponential decay for  $5 \lesssim t\lesssim  8$ and after that we find  a power-law regime for $9 \lesssim t\lesssim  15$. On the other hand, for $K=17$ only the exponential regime is observed in the interval  $4 \lesssim t\lesssim  7$  before the correlator function saturates for long times. The dashed black lines  in the main panel were obtained by fitting $|O_1(t)|\sim e^{-\gamma t}$. In the  inset we show the case of $K=6.6$ in log-log scale, and the black dashed line corresponds to a power-law fit $|O_1(t)|\sim t^\alpha$.}
\end{figure}

Let us now consider the behavior of the correlator $O_1$ for this map and different values of the kick strength $K$.  In Fig. \ref{fig:numeric_decay} we show a numerical evolution of the $O_1$ up to time $t=36$ (the time unit is a period) for  $K=6.6$ (circles) and for $K=17$ (squares). A Hilbert dimension of $D=5000$ was used in both plots. The four different regimes described in Fig. \ref{fig:analysis} can be clearly seen for $K=6.6$. The dashed black lines in the main panel correspond to an exponential fit $|O_1(t)|\sim e^{-\gamma t}$, while in the inset we show the case  $K=6.6$ in log-log scale to expose the power-law regime, and the (shifted) black dashed line marks the corresponding fit. After these regimes, $|O_1|$ slowly reaches the saturation value. In contrast, for $K=17$ (main panel Fig. \ref{fig:numeric_decay})   only the exponential regime between $t=2$ and $t=7$ can be observed before saturation.  We also can appreciate differences in the classical phase space structure in these cases. For instance, notice that for $K=6.6$, there are regular islands in this classical phase space, that are appreciable with the naked eye (see the bottom left panel of Fig.~\ref{fig:edf}). In contrast, for the $K=17$ case, in the bottom center panel of Fig.~\ref{fig:edf}, no regular structures are visible. In Sec.~\ref{sec:relationship} we will analyze this relation in more detail.  

\begin{figure}%[h!]
\includegraphics[width=.45\textwidth]{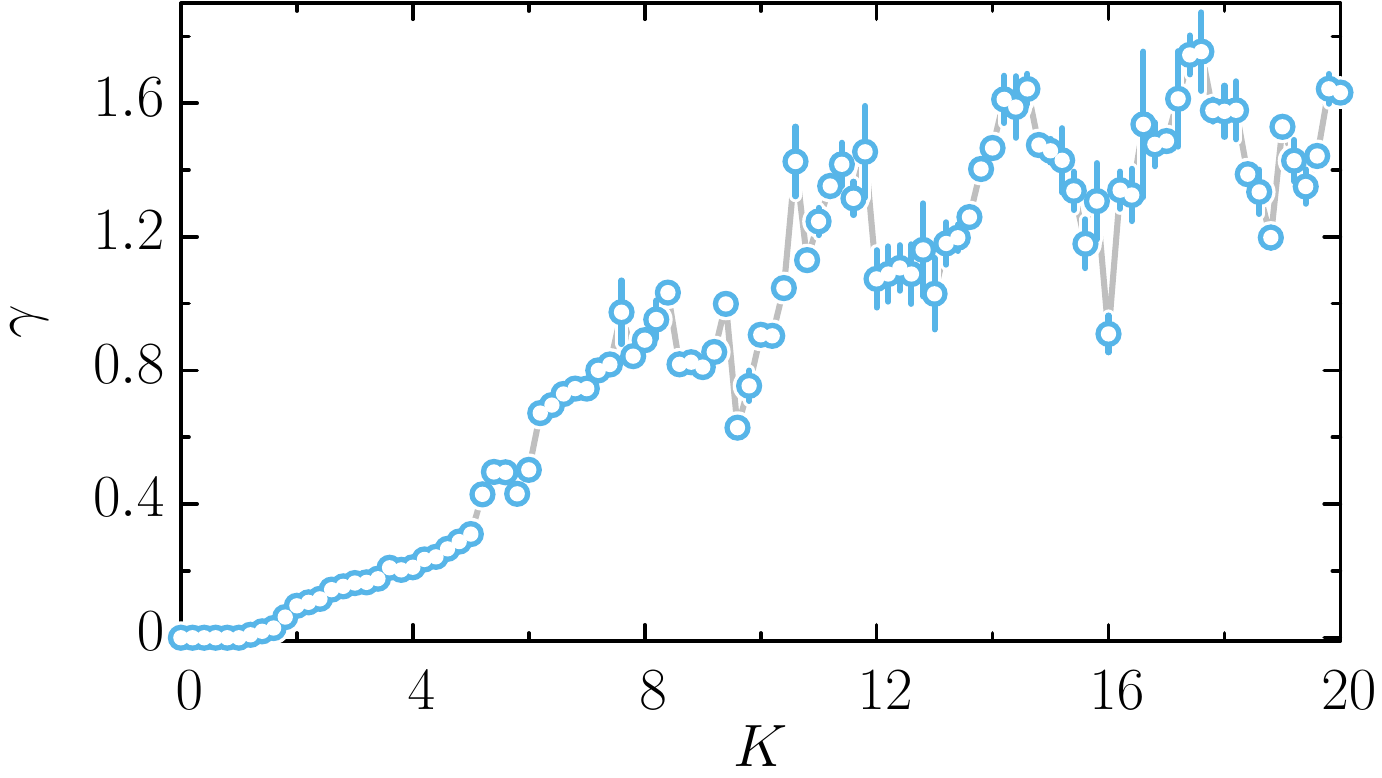} %{fig_gammadeK.pdf}
\caption{\label{fig:gammadeK} 
Decay rate $\gamma$ as a function of kicking strength $K$ obtained from fitting the  exponential  $|O_1(t)|\sim e^{-\gamma t}$ in suitably chosen time intervals.}
\end{figure}

In Fig.~\ref{fig:gammadeK} we show the decay rate $\gamma$ as a function of $K$ obtained by suitably fitting an exponential in the decay of $|O_1(t)|$. It can be clearly observed that for small $K$ the decay rate is negligible and a transition to strong exponential decay is observed for large $K$, consistent with chaotic behavior. The oscillating structure, marked by (almost) periodic dips, corresponds to the appearance or disappearance of small islands in the classical phase space.

\section{\label{sec:sepctrumPF} Numerical computation of the classical spectrum}
As we mentioned in Sec.~\ref{sec:O1 in maps} the Ruelle-Pollicott resonances are the point spectrum of the PF propagator. In the case of hyperbolic systems they establish the exponential rate of relaxation to the classical equilibrium density \cite{pollicott1985rate,ruelle1986,ruelle1987resonances}. 
For mixed systems, the appearance of regular islands in phase space generates sticking in these regions, affecting the relaxation rate, which, in this case, depends on the size of the regular regions. In these systems the Ruelle-Pollicott resonances are not well defined. Nevertheless, generalized resonances can be obtained using different numerical methods identifying stable eigenvalues inside the unit circle with eigenfunctions localized inside the chaotic regions \cite{Manderfeld2001,Weber2001,khodas2000relaxation,KhodasFishmanAgam}.

The task of computing the spectrum of the PF operator is not simple in any case (chaotic or mixed), and numerical approximation methods need to be considered. In this section we present the three methods we considered.
The first method \cite{blum2000leading, khodas2000relaxation}, that we called \textit{momentum-position method}, consists of expressing the Perron-Frobenius operator directly as 
\begin{equation}
    ( q',p' | \mathcal{L} |q,p )
    = \delta (p'-[p+\frac{K}{2\pi}\sin(2\pi q)] ) \, \delta\left(q' - [q+p']\right),
\end{equation}
in {the phase space basis  $\{|q,p)\}$, and
in terms of periodic $\delta$ functions on the circle.} In this way, the operator is unitary; thu, the spectrum is located in the unit circle. In order to study relaxation in these systems, one can coarse grained the dynamics by adding noise leading to a non-unitary operator. This can be done by replacing the $\delta$ function in the above expression by
\begin{equation}
    \delta(x) \to \sum_j \frac{1}{\pi s} \exp(-(x-j)^2/s),    
\end{equation}
where $s$ is a real parameter.
Then, the approximate spectrum is obtained by diagonalizing the operator in a reduced basis of $D$ vectors. Ideally, the resonances are obtained by diagonalizing the operator for a finite value of $s$ and then taking the limit of noise to zero.

The second method \cite{khodas2000relaxation,KhodasFishmanAgam,fishman2002relaxation} consists in expressing the PF operator in the Fourier-transformed phase space,  $( q,p | k,m ) = \exp(imq)\exp(ikp)/(2 \pi)$. In this case, noise is required for the convergence of the eigenvalues \cite{PhysRevLett.121.210601,garcia2003classical,khodas2000relaxation,fishman2002relaxation,nonnenmacher2003spectral}, especially for values of $K$ where the phase space is mixed.
Thus, after adding noise to $q$ with variance $\sigma^2$, the evolution operator can be written as:
\begin{equation}\label{eq:fishman}
\begin{split}
    (k,m | \mathcal{L^{(\sigma)}} |& k',m' ) \\
    &= J_{m-m'}(k'\,K)\exp(-\frac{\sigma^2}{2}m^2)\delta_{k-k',m}
\end{split}
\end{equation}
where $J_\nu (x)$ is the Bessel function of the first kind, $\sigma$ is the noise parameter. 
 Finally, we diagonalize a $D\times D$ matrix which represents a coarse-grained approximation of the PF operator. We will refer to this procedure as the \textit{Fourier method}. The advantage of this approach is that, the thicker the coarse-grained, the greater the importance of low modes associated with big structures of phase space. Namely, a low-dimensional matrix approximation can recover important features of the phase space, discounting the influence of small structures which usually are regular. The latter carries an advantage in terms of numerical efficiency. 

\par
\begin{figure}[ht]
\includegraphics[width=.49\textwidth]{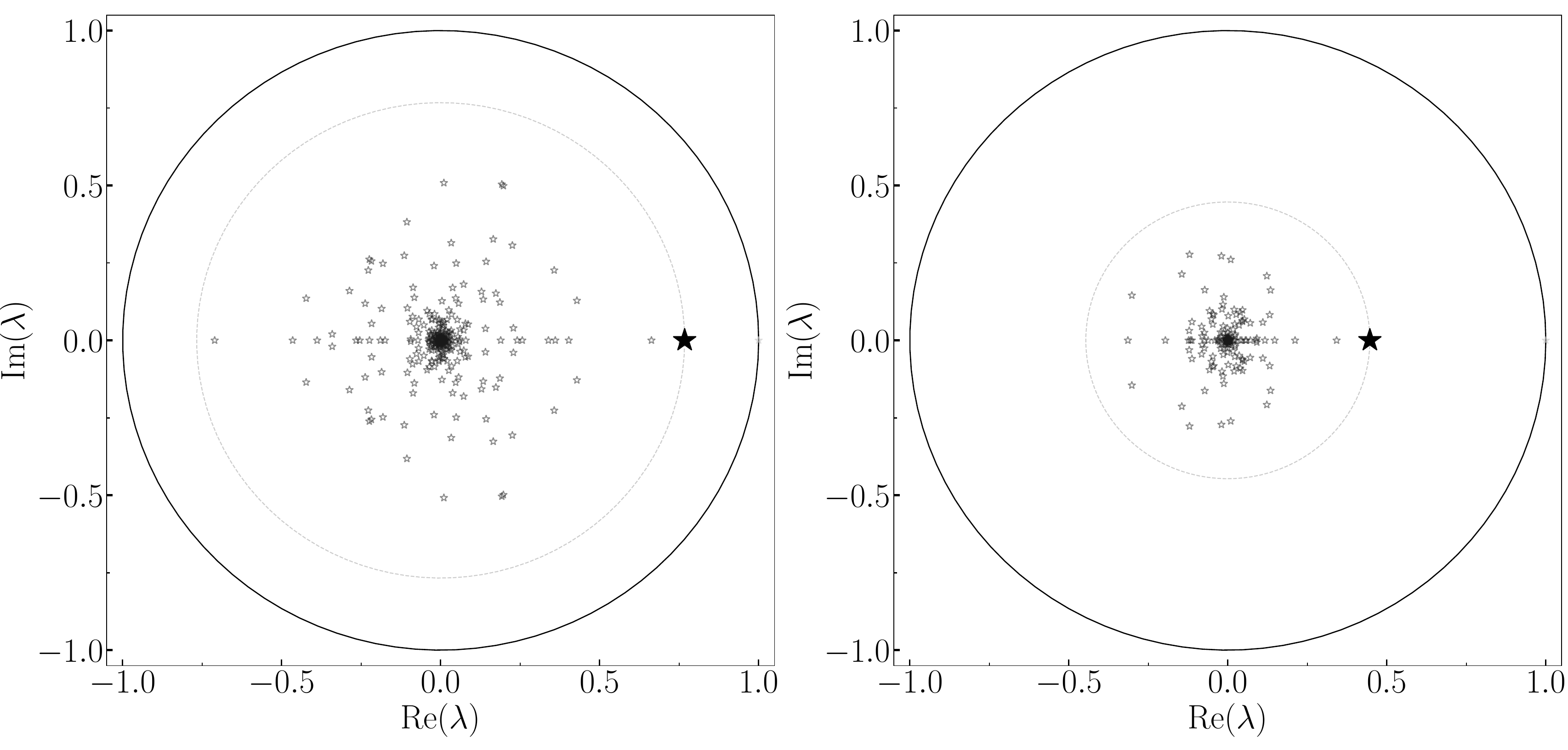}%{unit_circle_alt.pdf}
\caption{\label{fig:unit circle} Eigenvalues of the truncated Perron-Frobenius operator computed with the Fourier method. The left panel corresponds to $K=6.6$ and the right panel corresponds to $K=17$. In these cases we can see that there is a gap between the two leading eigenvalues such that $|\lambda|<1$, which lies on the real axis (big stars).  We see that, in general, the larger the chaos parameter $K$, the smaller the leading eigenvalues are. The matrix dimension is $D=30$ in both plots.}
\end{figure}

 In Fig. \ref{fig:unit circle} we exhibit an example of the spectrum for the truncated Perron-Frobenius operator for $K=6.6$ (left panel) and $K=17$ (right panel). The greater eigenvalue has modulus 1, and the next one is the leading eigenvalue $\lambda_1$, which we will associate with the decay of $|O_1|$. In this case, we observe that the modulus of the leading eigenvalue for $K=6.6$ is greater than the one for $K=17$. As we will show below, the smallest eigenvalues are associated with higher decay rates.

\par Finally, we have also  considered the Ulam method~\cite{ulam1960collection} to calculate the leading eigenvalues of the PF operator. This procedure consists of dividing the phase space into $N_d = M \times M$ cells and propagating $N_c$ trajectories on one map iteration from each cell $j$. Then the matrix $S_{ij}$ is defined by the relation $S_{ij} = N_{ij}/N_c$ where $N_{ij}$ is the number of trajectories moving from a cell $j$ to a cell $i$. By construction $\sum_i S_{ij} = 1$, and therefore, the matrix $S_{ij} \in \mathbb{R}^{N_d \times N_d}$ pertains to the class of PF operators (see \cite{meyer2000perron,pillai2005perron}) and can be considered as a discrete approximation of the PF operator of the continuous classical map \cite{frahm2010ulam}. However, we have observed that when there are regular regions present, this method is less accurate, so we favor the other two methods.

\section{\label{sec:relationship} Classical decay of $O_1$}
In Sec.~\ref{sec:standard} we showed examples of  
 the decay  $|O_1(t)|$,  for different values of the parameter $K$, as well as the dependence of the decay rate on the chaos parameter.   
Here we will show evidence that the  decay rate of $|O_1|$ shown in Fig.~\ref{fig:gammadeK} is closely related to $|\lambda_1|$  by  
\begin{equation}
    \gamma \approx -2\ln(|\lambda_1|). 
    \label{eq:gammalambda}
\end{equation}
We can stress that the left hand side of the equation is obtained from the quantum decay rate of $|O_1|$, while the term on the right hand side is a classical quantity. To this end we  used the methods described in the preceding section to  calculate the leading eigenvalue $|\lambda_1|$ (with modulus smaller than 1) of the Perron-Frobenius operator.
 For the momentum-position method we choose a noise value $s=0.001$ that is based on previous work \cite{blum2000leading}, while for the Fourier method the noise is $\sigma = 0.2$. 
 For the Ulam method, we have observed that the results, for the mixed dynamics case, do not fit  the ones obtained from the decay rates of $|O_1|$ (data not shown).

\begin{figure}%[h!]
\includegraphics[width=.45\textwidth]{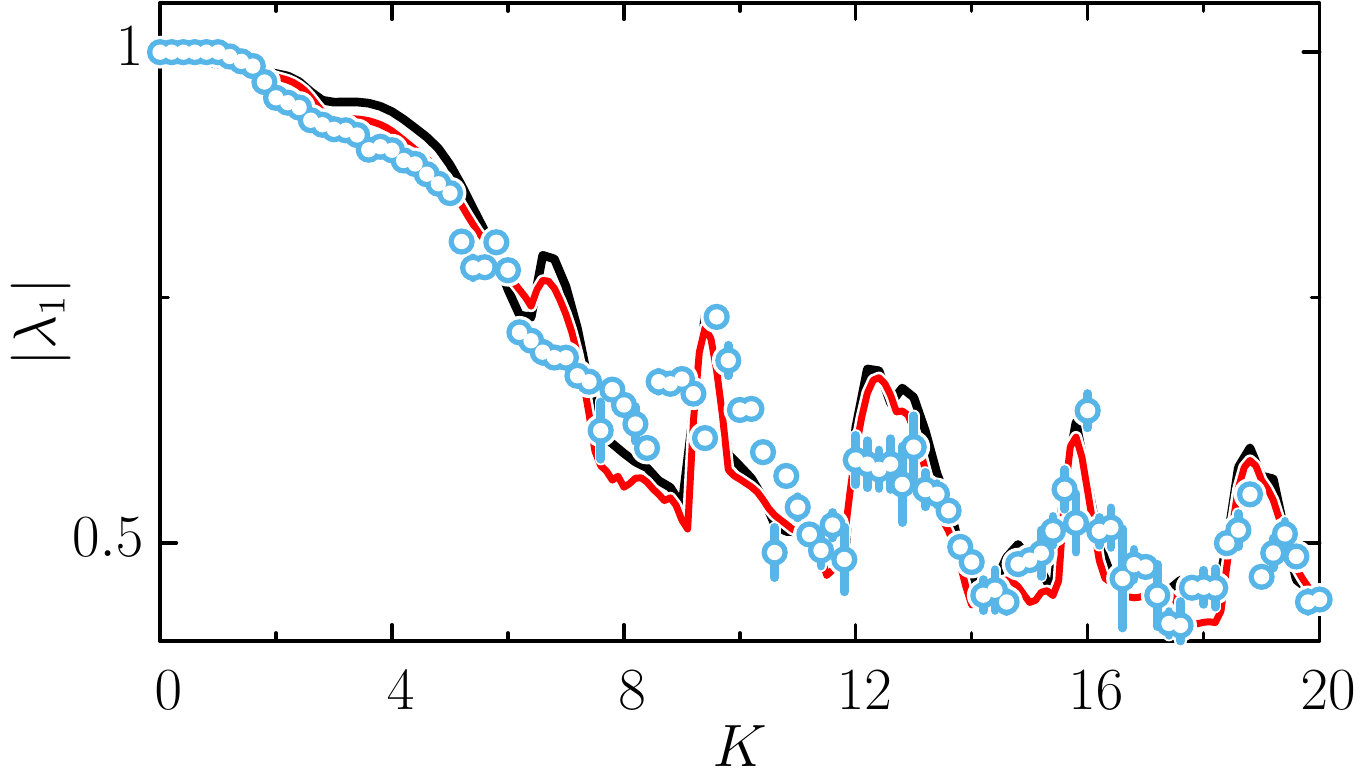}
\caption{\label{fig:resonances_and_decays} 
Absolute value $|\lambda_1|$ of the leading eigenvalue of the Perron-Frobenius operator as a function of $K$.
The black line corresponds to the momentum-position method,
and the red line corresponds to the Fourier method. These are compared to the values obtained from the decay rate of $|O_1|$ as $e^{-\gamma/2}$ (blue circles), the error bars are least squares errors due to the fit. 
For the Fourier method the dimension of the PF approximation was $D=30$ and the noise parameter $\sigma=0.2$, while  for the momentum-position method we used $D=90$ and $s=0.001$.}
\end{figure}

In Fig.~\ref{fig:resonances_and_decays} we show 
 the leading eigenvalues of the Perron-Frobenius operator $|\lambda_1|$ that, according to Eq.~\eqref{eq:gammalambda}, is equivalent to  the quantity $e^{-\gamma/2}$ obtained from the decay of $|O_1|$.   
In Fig.~\ref{fig:resonances_and_decays} we can see good agreement between the resonances obtained from both methods, and the data obtained from the fitted decay rate. From this  it is evident that $|\lambda_1|$, in general, decreases with the chaos parameter $K$. However, we also observe that for large values of $K$, even the system seems to be in a fully chaotic region, there are some bumps in the curve. {Below we show evidence that the four} ``bumps'', in $K\in[8,11]$, $K\in[12,14]$, $K\in[15,17]$ and in $K\in[18,20]$, are associated to the recovery of regularity in the phase space of the classical map. 

\begin{figure}[ht]
\includegraphics[width=.45\textwidth]{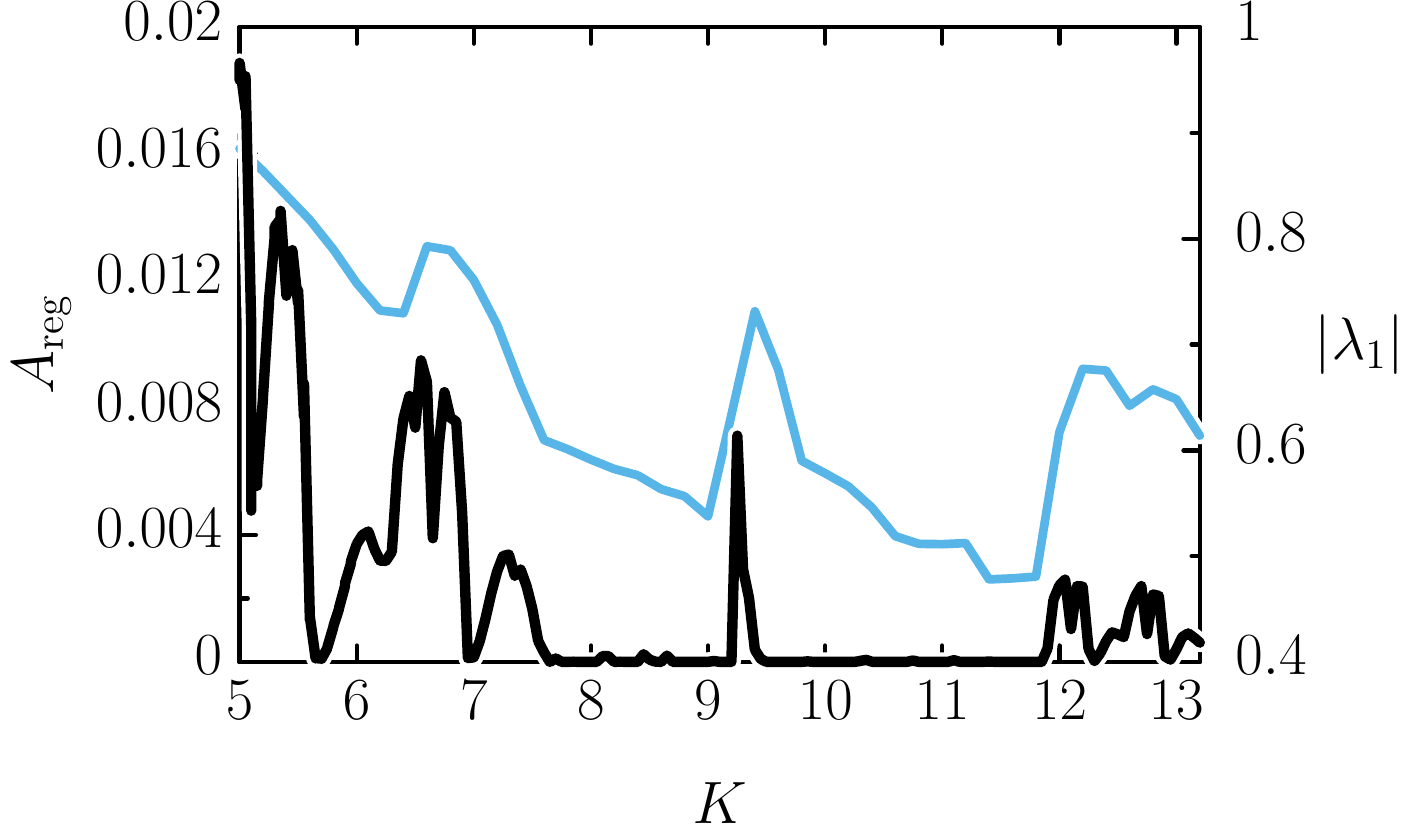}
\caption{\label{fig:hole} 
The black thick line (left $y$ axis) represents an estimation of the area of phase space occupied by regular regions for the  standard map as a function of $K$. The light blue line (right $y$ axis) corresponds to the calculation  of $|\lambda_1|$ using the momentum-position method. 
}
\end{figure}

A systematic way to show this is to estimate the area in phase space $A_{\rm reg}$ occupied by the regular islands as a function of the chaos parameter $K$.  To do so we randomly choose a large number $N_{\rm tot}$ of initial conditions. We evolve them with the map, but we introduce a hole in the chaotic region. By evolving for a large enough time, the only remaining points  $N_{\rm r}$ will be those that were initially inside the regular regions, which are disconnected from the chaotic ones. After a large number of time steps, and approximation of the regular area is 
\begin{equation}
 A_{\rm reg}\approx N_{\rm r}/N_{\rm tot}.   
\end{equation} 
If the whole phase space is chaotic, then $A_{\rm reg}\to 0$.
In Fig.~\ref{fig:hole} we show how the appearance of islands is manifested in the bumps of the black lines, and these are directly related to the bumps that appear in both the calculation of the resonance $|\lambda_1|$ and in the decay rate of $|O_1|$.
%%%%%%%%%%%%%%%%%%%%%%%%%%%%%%%%%%%%%%%%%%%%
%%%%%%%%%%%%%%%%%%%%%%%%%%%%%%%%%%%%%%%%%%%%

 Thus, our results suggest that the intermediate times of the quantum out-of-time-ordered correlator $O_1$ are governed  by classical magnitudes and structures. In this case, the eigenvalues of the PF operator play a  role similar to the Ruelle-Pollicott resonances in fully chaotic systems. At the same time, these eigenvalues inherit some characteristics of the phase space structure, as we showed by analyzing the revivals of regular islands in the chaotic region. 

\section{\label{sec:conclusions} Final remarks}
Although the OTOC has been around for quite some time, it  only recently attracted a lot of attention. The main motivation is related to understanding scrambling of quantum information. The other motivation, no less important, is associated with the characterization of the deep relation between chaos and scrambling. As such, there are classical quantities that can emerge from the time behavior of the OTOC. The Lyapunov exponent is the most widely known example because of the butterfly effect. However, the chaotic behavior is also characterized by relaxation and this is governed by the Ruelle-Pollicott resonances. In this work, we have extended the results obtained in \cite{PhysRevLett.121.210601} by showing that for a mixed system, generalized resonances dominate the approach to equilibrium (or to the asymptotic behavior) of the OTOC for a wide range of chaos parameter. The results were obtained for the quantum standard map, which is a kicked system, which depends on one parameter which allow us to cover regions ranging from fully integrable to fully chaotic.  In fact, we have also been able to show that the variations in the decay rates mimic very well the intricacies and variations of the resonances that reflect the appearance (and subsequent disappearance) of small regular islands for certain, quasi-periodic, values of the kicking strength. 

Unveiling the classical skeleton of quantum complex systems has, at least, two visible advantages. One is that the classical quantities determining some aspects of quantum time behavior may provide conceptually rich insight and, in some cases, may be easier to compute. Second, in the cases where there is no clear classical analog, they can provide powerful tools for semiclassical approaches \cite{prosen2004ruelle}.
We expect to make advances in that direction in future efforts. 

\begin{acknowledgments}
The authors thank M. Saraceno for interesting discussions. This study was (partially) supported by CONICET (Grant No. PIP 11220200100568CO), UBACyT (20020170100234BA) and ANPCyT (PICT-2020-SERIEA-01082 and PICT-2020-SERIEA-00740).  I.G.-M. received support from CNRS (France) through the International Research Project (IRP) COCSYS.
\end{acknowledgments}

\appendix

\section*{\label{app:convergence} Appendix}

\subsection*{Eigenvalues of Perron-Frobenius operator}

\begin{figure}
\includegraphics[width=0.45\textwidth]{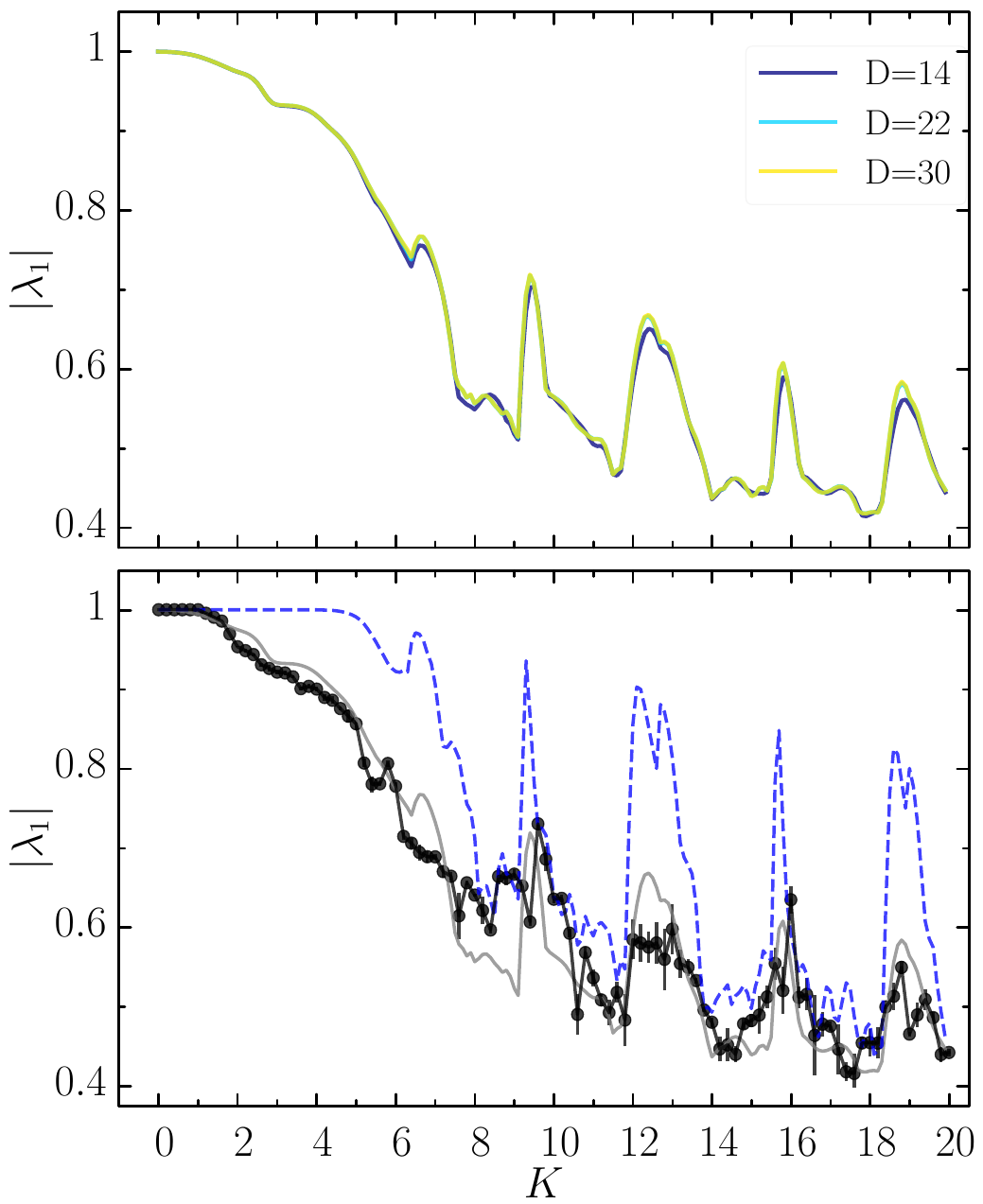} %{fig8app.pdf}
\caption{\label{fig:fishman} 
{Top:} Eigenvalues $|\lambda_1|$ of the PF operator computed with the Fourier method versus the chaos parameter for increasing values of $D$ and noise $\sigma = 0.2$.
{Bottom:} Eigenvalues $|\lambda_1|$ of the PF operator calculated with the Fourier method without noise (blue dashed lines) and $e^{-\gamma/2}$ obtained from decay rates $\gamma$ (black dots and solid line) versus the chaos parameter. The eigenvalues calculated with noise $\sigma = 0.2$ are shown by gray lines. The PF discrete matrix has dimension $D=30$.}
\end{figure}

 As we discussed in Sec.~\ref{sec:introduction}, the different decay rates  $\gamma$ of the correlator are related to the eigenvalue $\lambda_1$ of the Perron-Frobenius operator which has the greatest modulus such that $|\lambda_1|<1$ i.e., the leading eigenvalue inside the unit circle in the complex plane ($|\lambda_0|=1>|\lambda_1|>|\lambda_2|>...>0$).
 Therefore, in order to calculate these eigenvalues, the PF operator is approximated by a finite dimensional one of dimension $D$. Here, we show that, for fixed noise in position $\sigma = 0.2$, the eigenvalues of the Perron-Frobenius operator converge rapidly using the Fourier method. In Fig.~\ref{fig:fishman} we show $|\lambda_1|$ in terms of the chaos parameter for increasing matrix dimensions. 
It is clear that for $D \geq 22$  the eigenvalues associated to the different values of K converge to their asymptotic value. 
 As we mentioned in the main text, we need to add a given amount of noise to the calculation in order to converge to the value of $\lambda$ obtained from the decay of $O_1$. This can be observed in Fig.~\ref{fig:fishman}, where we show  $|\lambda_1|$ calculated with and without noise along with the corresponding $e^{-\gamma/2}$ obtained from the decay rate of $|O_1|$ [see Eq.~\eqref{eq:gammalambda}]. It can be seen that in the regions of $K$ where there are ``bumps'' in the decay rates (islands in the classical phase space), the eigenvalues $|\lambda_1|$ do not converge to the fitted values. For these regions, the addition of noise is essential.

\begin{figure}
\includegraphics[width=0.48\textwidth]{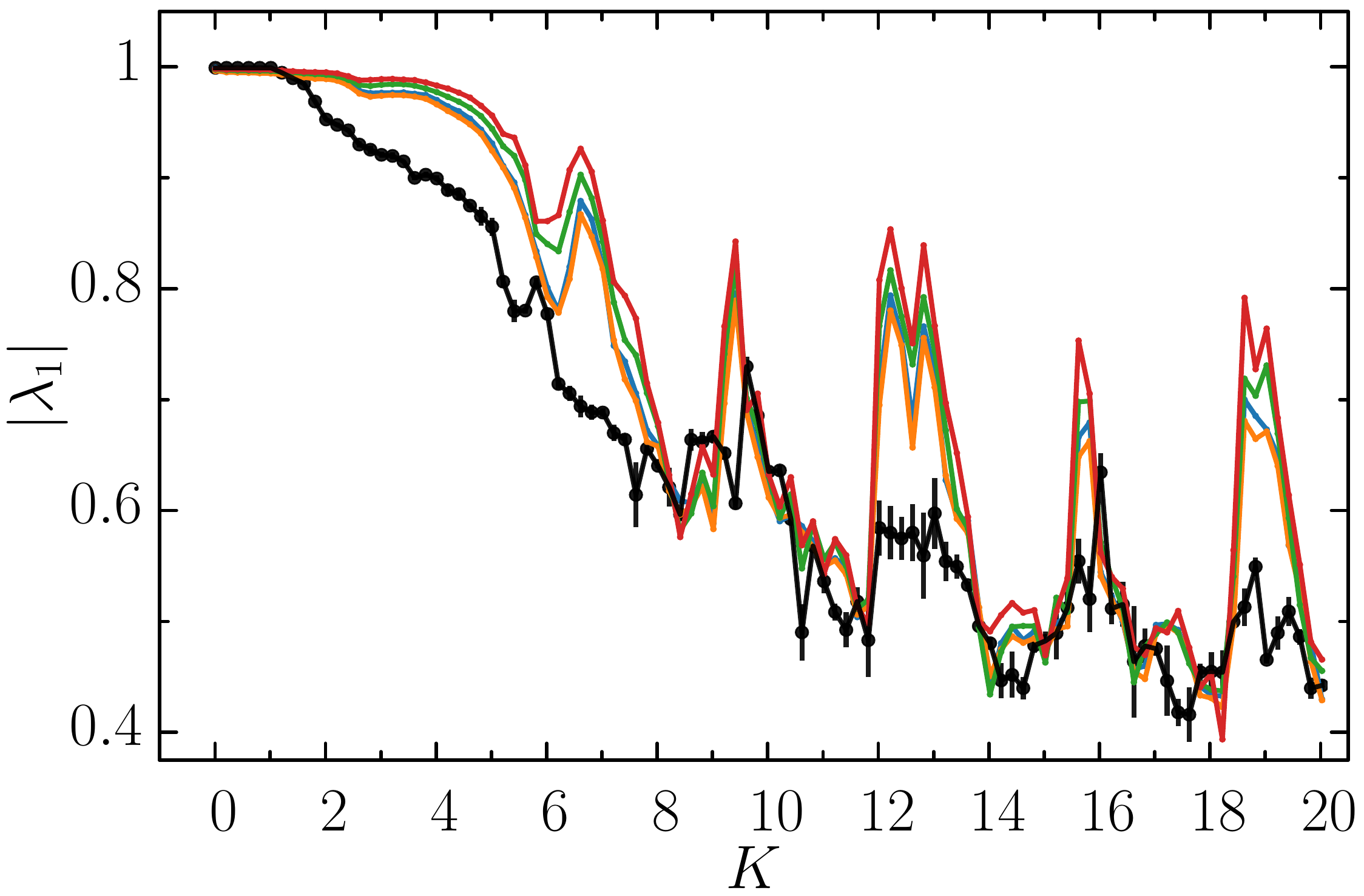}
\caption{\label{fig:Ulam} {Ulam eigenvalues versus $K$ for different matrix dimensions $D$. For these values, $D=30,40,50$ (orange, blue, and green lines respectively) we show the eigenvalues calculated with the addition of noise with normal distribution $\mathcal{N}(0,\hbar_{e\!f\!\!f})$ with $\hbar_{e\!f\!\!f} = 1/2\pi D$. We also show the eigenvalues for $D=30$ without noise (red line). }}
\end{figure}

{ In Fig.~\ref{fig:Ulam} we show $|\lambda_1|$ computed using the Ulam method for various matrix dimensions. 
This noise is inspired by a quantum coarse graining of the map. 
It can be seen that with or without noise the approximation of the eigenvalues fails mainly in the ``bumps'' associated with the recovery of regularity.
Although it can be observed that the values obtained from the Ulam method qualitatively reproduce  the bumps that correspond to the appearance of regular islands, they do not clearly reproduce the values of the decay rate of $|O_1|$ in these regions (as we have pointed out in the main text).}

%\bibliography{bibliography}% Produces the bibliography via BibTeX.
\providecommand{\noopsort}[1]{}\providecommand{\singleletter}[1]{#1}%

\end{document}